\documentstyle[12pt,psfig]{l-aa}

\begin{document}
\newcommand{\beq}{\begin{equation}}
\newcommand{\eeq}{\end{equation}}
\newcommand{\beqar}{\begin{eqnarray}}
\newcommand{\eeqar}{\end{eqnarray}}

\thesaurus{11(02.18.8;02.19.2;03.13.4;11.01.2;11.19.1)}
\title{Anisotropic illumination of AGN's accretion disk by a non thermal
  source . II General relativistic effects}
\author{P.O. Petrucci \and G. Henri}
\institute{Laboratoire d'Astrophysique, Observatoire de Grenoble,B.P 53X,
F38041 Grenoble Cedex, France}
\date{Received ??; accepted ??}
\maketitle
\markboth{P.O. Petrucci \& G. Henri: Anisotropic illumination of AGN's II}{}
\begin{abstract}
 In a previous paper (Henri \& Petrucci \cite{Hen*}, hereafter paper I),
 we have derived a new 
 model in order to explain the UV and X-ray emission of radio quiet
 AGNs. This model assumes that a point source of  
 relativistic 
 leptons ($e^+,e^-$) illuminates the accretion disk of the AGN by
 Inverse Compton process. This disk is supposed to be simply represented
 by a finite slab which radiates only the energy reprocessed from the
 hot source. The radiation field within the hot source region is
 therefore highly anisotropic, which strongly influences the Inverse
 Compton process. The different Eddington parameters characterizing the
 radiative balance of this system have been calculated self-consistently
 in the Newtonian case (paper I) giving a universal spectrum for a given
 inclination angle. In this paper, we take into account
 relativistic 
 effects by including the gravitational redshift, the Doppler boosting
 and the
 gravitational focusing due to the central supermassive black hole.
 This has the effect of modifying the radial temperature profile in the
 innermost region of the disk (at some gravitational radii). However, the
 spectrum is hardly different from that obtained in the Newtonian case,
 unless the hot source is very close to the black hole.
 These results are
 clearly different from standard accretion disk models where the
 gravitational energy is mainly 
 released in the vicinity of the black hole.

\end{abstract}

\section{Introduction}
 It is now generally agreed that the engine of high power emission in AGNs
 is a supermassive black hole of $10^{6}-10^9\,M_{\sun}$, accreting matter
 from a surrounding accretion disk (Shakura \& Sunyaev \cite{Sha73}; Rees
 \cite{Ree84}). Besides,  
 several satellites observations of radio quiet Seyfert galaxies have
 allowed to obtain an average high energy (X-ray/$\gamma$-ray) spectrum,
 better reproduced by a complex superposition of a primary power law, a
 reflected component from a cold thick gas, a fluorescent iron K line and
 an absorption by a hot medium (Pounds et al. \cite{Pou90}). On the other hand,
 Clavel et al. (\cite{Cla92}) have shown a close simultaneity between UV
 and optical 
 variations of some Seyfert galaxies, which cannot be reproduced by
 standard accretion disk models. Rather, these results are better
 explained if the UV-optical emission comes from the reprocessing of hard
 radiation emitted by a hot source above the disk.
 In paper I, we have proposed a new model involving a point source of 
 relativistic leptons (the hot source) emitting hard radiation by Inverse
 Compton (IC) process on soft photons produced by the accretion disk. The
 disk itself radiates only through the reprocessing of the hard
 radiation impinging on it.
 Such a geometry is highly
 anisotropic, which 
 takes a real importance in the computation of IC process (Ghisellini et
 al. \cite{Ghi91} 
 ; paper I). Paper I dealt only with the Newtonian
 case and did not include
 the relativistic effects: these are first, the Doppler shift due to the
 rotation of the 
 disk; second the gravitational shift, undergone by photons which follow a null
 geodesic, either from the disk to the hot source and inversely, or from
 the AGN to the observer at infinity; and third, the 
 gravitational focusing, most important for rays of light skimming the
 black hole. Thus, the subject of this paper
 (paper II) is to extend this simple model to the general relativistic
 formulation appropriate to a Kerr black hole.
 The organization is as follows. We will establish, in
 section 2, the general equations which govern the radiative balance
 between the hot source and the accretion disk in an axisymmetric
 gravitational field.  We will study the case of a rotating black hole in
 the Kerr metrics in section 3. In section 4, we will finally obtain the power
 spectra emitted by this model for different values of the inclination
 angles and the height of the hot source, and
 conclude on the importance of the gravitational effects
 on the overall spectrum.
 \begin{figure*}
  \psfig{width=17cm,height=4cm,angle=-90,file=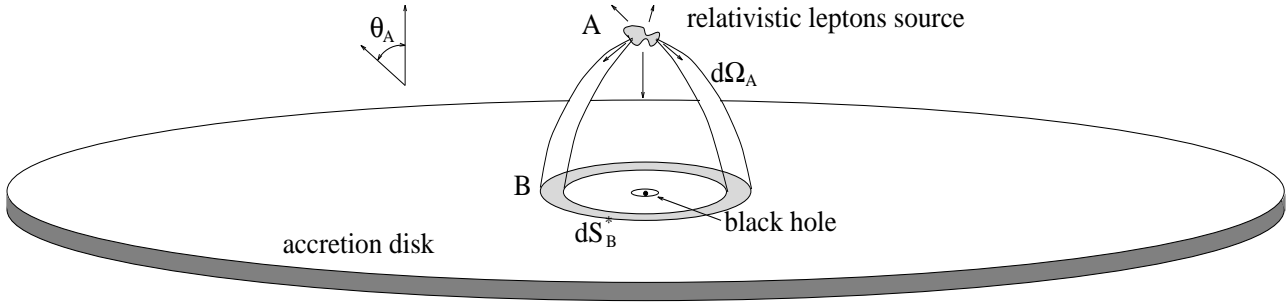}
  \caption[]{The general picture of the model. We have also drawn the
    trajectory 
    of a beam of photons  emitted by the hot source in a solid angle
    $d\Omega_A$ and absorbed by a surface ring $dS_B$ on the
    disk. The letters A and B refer to the indices define in part
    (\ref{def})}\label{figgen}   
 \end{figure*}

\section{Relativistic energy balance equations}
 The geometry of the
 system, formed by the hot source and the accretion disk, is sketched in
 Fig. \ref{figgen}. The energy balance can be solved self-consistently,
 for a given disk emissivity law. Actually, the
 emission of the disk is entirely controlled by the hard
 radiation angular distribution, which is at turn determined by the disk
 emissivity through the anisotropic IC process.
 In paper I, we solved 
 the Newtonian case of this radiative balance by solving a system of 3
 equations between the 3 Eddington parameters characterizing the photon
 field. When relativistic effects are taken into account, the same principle
 can be used, but with some modifications. First, the photons do not
 follow any more straight trajectories, but geodesics whose equations
 must be deduced from the metrics. Second, one must take care of
 gravitational and Doppler shifts between the hot source, the rotating
 accretion disk and the observer at infinity.

\subsection{The different frames}
 \label{def}
 The Kerr metrics describes the exterior metrics of an stationary
 axisymmetric gravitational field around a rotating body. It is
 completely specified by its total mass $M$ and angular momentum per unit
 mass $a$. The line element can be written in Boyer-Lindquist
 coordinates as follows (we use 
 convenient units such that $c=G=1$; then the unit length is
 $GM/c^2=R_g/2=M$, $R_g$ being the Schwarzschild radius):
 \beq 
  \label{kerrmetric}
  ds^2=-e^{2\nu}dt^2+e^{2\psi}(d\phi-\omega dt)^2+\Delta^{-1}\Sigma dr^2+
  \Sigma d\theta^2 
 \eeq 
 where 
 \beqar
  A &=& (r^2+a^2)^2-a^2\Delta\sin^2\theta \\ 
  \Delta &=& r^2+a^2-2Mr \\ 
  \Sigma &=& r^2+a^2\cos^2\theta \\
  e^{2\nu} &=& A^{-1}\Delta\Sigma \\ 
  e^{2\psi} &=& A\Sigma^{-1}\sin^2\theta \\
  w &=& 2aMrA^{-1}. 
 \eeqar 
 Yet, as shown by Bardeen et al. (\cite{Bar72}), physics is not simple in
 the Boyer-Lindquist coordinate frame. First, the dragging of inertial
 frame becomes so severe as we approach the Kerr black hole, that the
 line element $ds^2$ goes time-like. Second, the metrics is non-diagonal
 which introduces algebraic complexities. For those reasons, Bardeen
 introduced the locally non-rotating frames (LNRF, Bardeen et
 al. \cite{Bar72}) to 
 cancel out, as much as possible, the ``frame dragging'' effects of the
 hole rotation. They are linked to observers whose world lines are
 $r=constant$, $\theta=constant$ and $\phi=\omega t+ constant$. We will
 used the same method here. 
 We defined the set of frames $\Re$ as these LNRF. For a Schwarzschild black
 hole, they correspond simply to 
 the curvature coordinates frame $(r,\ \theta,\ \phi)$. 
 We also define the set of frames $\Re^*$ which locally rotate with the
 disk. All terms computed in these frames are labeled with a star.\\
 The quantities expressed at the hot source location ($\theta=0,\ r=Z_0$)
 are indexed with an A. They are indexed with a B when they are computed
 on the surface of the disk ($\theta=\frac{\pi}{2}$). For example, a
 differential elementary surface of the disk, expressed in the rotating frame
 $\Re^*$, will be noted $dS^*_B$ (cf. Fig. \ref{figgen}).
 The gravitational shift between any point P and Q is noted
 $(1+z)_{PQ}=
 \displaystyle\frac{(1+z_{P\infty})^{PQ}}{(1+z_{Q\infty})^{PQ}}$, with
 $(1+z_{P\infty})^{PQ}\equiv\lambda_o-\lambda_e$, 
 $\lambda_e$ being the emitted wavelength of a    
 photon and $\lambda_o$ the observed wavelength at infinity along the
 geodesic connecting $P$ and $Q$.

\subsection{Computation of the specific intensity}
 The radiative balance between the energy radiated by the disk and that
 radiated by the hot source of relativistic leptons gives the 
 relation:
 \beqar
   \label{eq1}
   F^*(r_B) &=& \frac{dP^*_B}{dS^*_B} \nonumber \\ &=&
   (1+z)^{2}_{B^*A}\left(\frac{dP_A}{{d\Omega}_A}
     {{d\Omega}_A}\right)\frac{1}{dS^*_B}. 
 \eeqar 
 Here, $F^*(r_B)$ is the
 flux emitted in the frame rotating with the disk, by the surface ring
 which radius is in the range [$r_B$,$r_B+dr_B$], and we use the fact
 that the ratio $\displaystyle\frac{dP_Q}{(1+z_{Q\infty})^2}$ (where $dP_Q$
 is the 
 radiative power, measured in $Q$, released by the hot source) is a
 relativistic  invariant. 
 Using, now, the covariance of the space-time quadrivolume between the 2
 inertial frames $\Re$ and $\Re^*$, we obtain:
 \beq
   \label{eq3}
       \underbrace{{dS^*_B}{dt^*_B}{dh^*_B}}_{\rm{in\ }\Re^*}=
        \underbrace{{dS_B}{dt_B}{dh_B}}_{\rm{in\
            }\Re}
 \end{equation}
 where $dh^*_B$ and $dh_B$ are the elementary space intervals in the Z
 direction. Since there is no motion along this direction, $dh^*_B=dh_B$ and
 thus , combining Eqs. (\ref{eq1}) and (\ref{eq3}), one gets:
 \beqar
  F^*(r_B) &=& \left(\frac{dP}{{d\Omega}_A}{{d\Omega}_A}\right)
  \frac{1}{dS_B}\frac{dt^*_B}{dt_B}(1+z)^2_{B^*A} \nonumber\\ 
           &=& \frac{dP}{{d\Omega}_A}\frac{{d\Omega}_A}{dS_B}
           (1+z)_{BB^*}
           (1+z)^2_{B^*A}. \label{eqtot}
 \eeqar
 We suppose the disk to radiate isotropically like a blackbody at the
 temperature $T_{\rm eff}$. So, one gets also:
 \beqar
  F^*(r_B) &=& \pi I^*_B(r_B) \label{eqI1}\\
           &=& \sigma T_{\rm eff}^4(r_B).    \label{eqBB}
 \eeqar
 The ratio $\displaystyle\frac{dP}{d\Omega_A}$ of the power
 emitted by the hot source by solid angle unit, is derived 
 in Eq. (48) of paper I . One gets (with $\mu_A=cos\theta_A$, cf.
 Fig. \ref{figgen}):  
 \begin{equation} 
 \frac{dP}{d\Omega_A}=\frac{3L_t}{32\pi
   J_A}[(3J_A-K_A)-4H_A\mu_A+(3K_A-J_A)\mu_A^2] 
 \label{Xspec}
 \end{equation} 
 where $J_A,\ H_A$ and $K_A$ are the three Eddington parameters defined by:
 \begin{eqnarray}
   J_A &=& \frac{1}{2}\int_{-1}^{1}I(\mu_A)d\mu_A \nonumber\\
   H_A &=& \frac{1}{2}\int_{-1}^{1}I(\mu_A)\mu_Ad\mu_A \label{Edd}\\
   K_A &=& \frac{1}{2}\int_{-1}^{1}I(\mu_A)\mu_A^2d\mu_A. \nonumber
 \end{eqnarray}
 The  gravitational shifts
 $(1+z)_{B^*A}$ and $(1+z)_{BB^*}$ and the Jacobian
 $\displaystyle\frac{{d\Omega}_A}{dS_B}$ of Eq. (\ref{eqtot}) will be
 developed in the next section.
 \noindent
 We deduce from Eqs. (\ref{eqtot}), (\ref{eqI1}) and (\ref{Xspec})
 the general 
 expression of the specific intensity of the radiation emitted by the
 disk, in the rotating frame $\Re^*$:
 \beqar
        \label{eqI}
        I^*_B(r_B) &=& (1+z)^2_{B^*A}(1+z)_{BB*}
        \frac{d\Omega_A}{dS_B}
        \frac{3L_t}{32\pi^2J_A}\times \nonumber \\
        & & [3J_A-K_A-4H_A\mu_A+(3K_A-J_A)\mu^2_A]. 
 \eeqar

\subsection{Computation of Eddington parameters}
\label{Eddcomp}
 Now, we can find a system of 3 equations between the 3 Eddington
 parameters $J_A$, $H_A$ and $K_A$ characterizing the radiation field near
 the hot source. Using Eqs. (\ref{Edd}) and (\ref{eqI}), and the fact that
 $\displaystyle\frac{I}{\nu^4}$ is a relativistic invariant (Liouville's
 theorem), one obtains:
 \beqar
   J_A &=& \frac{1}{2}\int_{-1}^{1}(1+z)^{-4}_{B^*A}I^*_Bd\mu_A \nonumber \\
       &=& \frac{1}{2}\int_{-1}^{1}(1+z)^{-2}_{B^*A}(1+z)_{BB*}
       \frac {d\Omega_A}{dS_B}\times  \\
   & & \frac{3L_t}{32\pi^2 J_A}[3J_A-K_A-4H_A\mu_A+
   (3K_A-J_A)\mu^2_A]d\mu_A. \nonumber
 \eeqar
 If we define, now, the following parameters:
 \beqar
   \xi_A &=& \frac{3L_t}{32\pi^2Z_0^2 J_A} \\ 
    G_n  &=& \frac{1}{2}\int_{-1}^{1}(1+z)_{B^*A}^{-2}(1+z)_{BB*}
         \frac{d\Omega_A}{dS_B}Z_0^2\mu_A^nd\mu_A,
 \label{z}
 \eeqar
 we can rewrite the expression of $J_A$, and those of the second and
 third Eddington moment $H_A$ and $K_A$ in the same way, in order to
 obtain the following linear system:
 \beq
 \left\{
 \begin{array}{lll}
   J_A &=& \xi_A[J_A(3G_0-G_2)-4H_AG_1+K_A(3G_2-G_0)]\\
       & & \\
   H_A &=& \xi_A[J_A(3G_1-G_3)-4H_AG_2+K_A(3G_3-G_1)]\\ 
       & & \\
   K_A &=& \xi_A[J_A(3G_2-G_4)-4H_AG_3+K_A(3G_4-G_2)].
 \label{linsys}
 \end{array}
 \right.
 \eeq
 We will find the values of $\displaystyle\eta_A=\frac{H_A}{J_A}$ and
 $\displaystyle\chi_A=\frac{K_A}{J_A}$ by making the determinant of this
 system vanish, that is by solving a cubic equation in
 $\xi_A$. The only physical constraints on the choice of $\xi_A$ that we
 have to respect are $\eta_A\leq 1$ and $\chi_A\leq 1$. Note that the
 Newtonian case of an infinite disk of paper I can be recovered by setting
 $G_n=\displaystyle\frac{1}{2(n+4)}$. In the general case, we need to
 compute the values of the 
 variables $G_n$, that is, to express the gravitational shifts
 $(1+z)_{B^*A}$ and $(1+z)_{BB*}$, and the ratio
 $\displaystyle\frac{d\Omega_A}{dS_B}$.
 This will be the subject of
 section 3 where we use the Kerr metrics to calculate explicitly these
 coefficients.

 \section{Computation in the Kerr geometry}
 The photons follow null geodesics either between the disk and the hot
 source, or between the disk/hot source and the observer at infinity. We
 recall the general expressions of the momentum along a null geodesic
 (Carter \cite{Car68}, Cunningham \cite{Cun75}):
 \beqar
  p_t &=& -E\\
  p_\phi &=& E\lambda\\
  p_r &=& \pm EV_r^{1/2}\Delta^{-1} \\
  p_\theta &=& \pm EV_\theta^{1/2}\Sigma^{-1} 
 \eeqar
 with
 \beqar
   V_\theta &=&
   q^2-(\frac{\lambda^2}{\sin^2\theta}-a^2)\cos^2\theta \\
   & & \nonumber \\
   V_r &=& (r^2+a^2-a\lambda)^2- \nonumber \\
   & & (r^2+a^2-2r)((\lambda-a)^2+q^2). 
 \eeqar 
 $E,\ q,\ \lambda$ are constants of motion: $E$ is the
 energy-at-infinity and $\lambda$ and $q$ are closely related to the angular
 momentum. For geodesics intersecting the Z axis, one has $\lambda=0$,
 which is the case for every photon coming from or reaching the hot source. 

 \subsection{The gravitational shifts}
 \label{kerrsourcespec}
 First, we need the expressions of the gravitational shifts of
 Eq. (\ref{z}). Since the hot source A is at rest, one gets:
 \beq
 (1+z_{A\infty})^{B^*A}=e^{\nu_A}=\left(1-\frac{2MZ_0}{Z_0^2+a^2}\right)^{1/2}.
 \eeq
 The shift $(1+z_{B^*\infty})^{B^*A}$ between a point $B^*$ rotating with
 the disk and the
 infinity, is given by Cunningham (\cite{Cun75}) (with $\lambda =0$ since
 the geodesic 
 crosses the hot source): 
 \beq
  (1+z_{B^*\infty})^{B^*A}=e^{\nu_B}(1-V_e^2)^{1/2}
 \eeq
  where $V_e$ is the velocity of the disk in the locally non-rotating
  frame $\Re$, which can be express as a function of the coordinate
  angular velocity of 
  the disk $\Omega_e$ (Cunningham \& Bardeen \cite{Cun73}): 
 \beqar 
  V_e &=& (\Omega_e-\omega)e^{\psi-\nu}\\
  \Omega_e &=& M^{1/2}(r^{3/2}+a)^{-1}. 
 \eeqar
 We thus obtain the following expression for the gravitational shift
 between $A$ and $B^*$:
 \beq
   (1+z)_{B^*A} = \frac{e^{\nu_A}}{e^{\nu_B}(1-V_e^2)^{1/2}}.
 \eeq
 The shift between $B$ and $B^*$ is deduced from Lorentz transformation between
 the 2 inertial frames $\Re$ and $\Re^*$, that is:
 \beq
   (1+z)_{BB^*} = (1-V_e^2)^{1/2} \label{eqzBB}.
 \eeq

 \subsection{Computation of $\displaystyle\frac{d\Omega_A}{dS_B}$}
 The disk surface element $dS_B$ contained between $r_B$ and $r_B + dr_B$
 is calculated in Appendix \ref{append1}:
 \beq
   dS_B=2\pi A^{1/2}\Delta^{-1/2}dr_B.
 \eeq
 Thus, we obtain:
 \beq
  \frac{d\Omega_A}{dS_B}=A^{-1/2}{\Delta^{1/2}}\frac{d\mu_A}{dr_B}.
 \eeq
 The derivative $\displaystyle\frac{d\mu_A}{dr_B}$ is computed numerically by
 integrating the equation of motion between the hot source and the disk,
 for a grid of initial values of $\mu_A$. The equation of motion has
 been obtained by Carter (\cite{Car68}) taking full advantage of the separation
 of variables: 
 \beq
   \int_{0}^{\pi/2}\frac{d\theta}{\sqrt{V_\theta}}=
   \int_{Z_0}^{r_B}\frac{dr}{\sqrt{V_r}}. 
   \label{eqtmotion}
 \eeq   
 The signs of $V_r^{1/2}$ and $V_\theta^{1/2}$ are always the same as the signs
 of $dr$ and $d\theta$, respectively. In this case, $d\theta$ is always
 positive (we do not take into account geodesics spinning round the black
 hole). Only $dr$ can change its sign at a turning point in $r$. The
 constant of motion $\lambda$ and $q$ must be taken such that, at the 
 starting point A, one has:
 \beqar
  p_\phi &=& 0\\
  p_rp^r &=& \mu_A^2 p_tp^t.
 \eeqar  
 This gives:
 \beqar
  \lambda &=& 0\\
  q &=& \left[(1-\mu_A^2)(Z_0^2+a^2)(Z_0^2+a^2-2MZ_0)^{-1}-a^2\right]^{1/2}.
 \eeqar
 Equation (\ref{eqtmotion}) is then solved with respect to $r_B$, for a
 given $\mu_A$.   
 Once all the coefficients $G_n$ are computed, the linear system
 (\ref{linsys}) can be solved, by making its determinant vanish. One can
 extract the values of $\eta_A$ and $\chi_A$ and compute the radial
  effective temperature distribution $T_{\rm eff}(r_B)$ by means of Eqs. 
 (\ref{eqI1}), (\ref{eqBB}) and (\ref{eqI}).

 \subsection{Disk emission spectrum}
 The power carried to the observer by the photons emitted by a surface
 element of the disk, will be the
 product of its observed solid angle and specific intensity. Using again
 the Liouville's theorem to relate the observed power to the emitted
 specific intensity $I_{B^*}$, measured in the rest frame of the emitter,
 we obtain:
 \beq
        \label{eq26}
        dP_{\nu_o}^{disk}=(1+z_{B^*\infty})^{-3}{I_{\nu_e}}_{B^*}d\Omega
 \eeq                   
 where $(1+z_{B^*\infty})$ is the redshift between the disk and the
 observer at infinity. Here again, we are only interested in the ``direct''
 geodesics and do not compute photon trajectories
 crossing several times the equatorial plane between the black hole and
 the observer. If 
 we suppose that the disk radiates like a 
 black-body, the specific intensity ${I_{\nu_e}}_B^*$ is simply the Planck
 function $B_{\nu_e}(T_{\rm eff}(r_B))$.
 \begin{figure}
        \psfig{angle=-90,width=\columnwidth,file=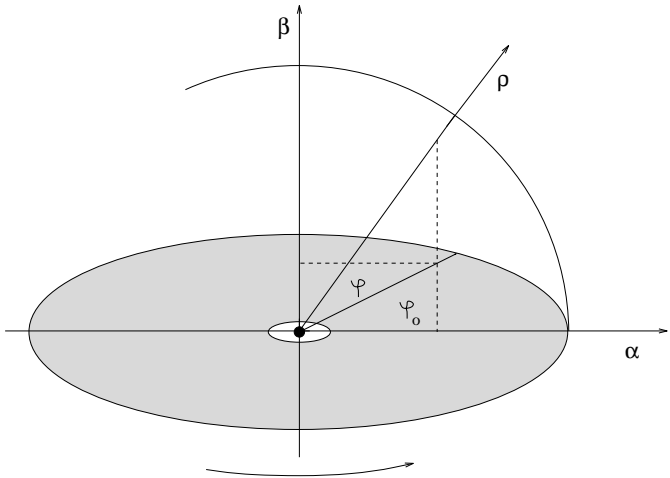}
        \caption[]{The impact parameters $\alpha, \beta$ in the plane of
          the sky deduced from the polar coordinates ($\rho,\varphi$). If
          $\theta_{\mbox{\scriptsize{0}}}$ is the inclination angle of
          the disk, one have the relation
          $\tan\varphi_{\mbox{\scriptsize{0}}}
          =\cos\theta_{\mbox{\scriptsize{0}}}\tan\varphi$}\label{skyproj}
 \end{figure}
 We spot the apparent position of a point P of the disk by
 its coordinates $(\alpha,\beta)$ in the plane of the sky. A couple
 $(\alpha,\beta)$  
 represents the coordinates of the impact parameter of the null
 geodesic between the disk and the observer. They are measured relative
 to the direction of the center of the black hole, in the sense of the
 angular momentum (see Fig. \ref{skyproj}).
 Once again, we use the Carter's formalism to compute the geodesic
 between the disk and the observer. We can simply expressed $\lambda$ and
 $q$ as a function of the impact parameters $\alpha,\beta$
 (Cunningham \& Bardeen \cite{Cun73}):
 \beqar
  \lambda &=& -\alpha\sin\theta_0 \label{lambda}\\
  q &=& (\beta^2-a^2\cos^2\theta_0+\alpha^2\cos^2\theta_0)^{1/2}.
  \label{q} 
 \eeqar
 Here, $\theta_0$ is the inclination angle of the accretion disk. The
 radius $r_B$ of the 
 emitting point of the accretion disk is then calculated by solving the
 new equation of motion: 
 \beq
   \int_{\pi/2}^{\theta_o}\frac{d\theta}{\sqrt{V_\theta}}=
   \int_{r_B}^{\infty}\frac{dr}{\sqrt{V_r}}. 
   \label{eqtmotion2}
 \eeq   
 Finally, we can expressed
 the gravitational redshift between a point of the accretion disk and an
 observer at infinity, needed in Eq. (\ref{eq26}),
 as follows (Cunningham \& Bardeen \cite{Cun73}): 
 \beq 
  (1+z)_{B^*\infty}=e^{-\nu}(1-V_e^2)^{-\frac{1}{2}}(1-\Omega_e\lambda). 
 \eeq 
 The total spectrum is computed by integrating Eq. (\ref{eq26}) over
 the disk surface. The grid in $(\alpha,\beta)$ is obtained from a
 elliptic polar coordinates sampling (Fig. \ref{skyproj}). The polar
 angles $\varphi$ are regularly spaced whereas 
 we use a logarithmic sampling of polar radius $\rho$. The Cartesian
 coordinates ($\alpha,\beta$) are then deduced by the following formulae:
 \beqar
  \alpha &=& \rho\cos\varphi \nonumber\\
  \beta &=& \rho\sin\varphi\cos\theta_0.
 \eeqar

 \section{Results and Discussion}
\subsection{The set of parameters}
 The method described above have been used to obtain spectra emitted by
 the accretion disk and the hot source. In the Newtonian case, as shown
 in Paper I, the disk
 emission depends only on the total luminosity $L_t$ and the height $Z_0$
 of the hot source above the disk. Furthermore, one finds a universal
 spectrum as a function of a reduced frequency $\nu/\nu_c$ and reduced
 luminosity $\nu F_{\nu}/L_t$ where
 \beq
  \nu_c = \frac{k_B}{h}\left(\frac{3L_t}{32\pi Z_0^2\sigma}\right)^{1/4},
  \label{nuc}
 \eeq
 corresponding to the characteristic temperature 
 \beq
  T_c = \frac{h}{k_B}\nu_c.
 \eeq
 In the relativistic calculations, one must also specify the mass $M$ and
 the angular momentum by unit mass $a$ of the black hole. Actually, the
 disk emission in reduced units depends only on $a$ and $Z_0/M$. However,
 one needs a value of $\nu_c$ comparable to the observations, i.e. 
 about $10\,eV$. As an example, for $L_t=10^{45}\,erg.s^{-1}$ and
 $M=5\,10^6M_{\sun}$, one gets 
 $\nu_c=85\displaystyle\left(\frac{M}{Z_0}\right)^{1/2}$eV that is
 $Z_0/M$ about $70$.   

 The high energy spectrum depends also on the relativistic particle
 distribution, which was taken as a exponentially cut-off power law
 (cf. paper I):
 \beq
  f(\gamma)\propto\gamma^{-s}\exp(-\frac{\gamma}{\gamma_0}).
 \eeq
 Thus, one needs also to specify the spectral index $s$, and the cut-off
 Lorentz factor $\gamma_0$ or equivalently the high energy cut-off
 frequency $\nu_0$. Again, the total spectrum is universal for a given
 value of $a$, $Z_0/M$, $\displaystyle\frac{\nu_0}{\nu_c}$ and
 $s$. The OSSE/SIGMA 
 observations favor the values $h\nu_0\simeq 100\,keV$ and $s\simeq 3$. We
 have kept these values for all simulations.

\subsection{Angular distribution of radiation}
 \label{paragraphangular}
 As already mentioned, the angular distribution of high energy radiation
 is entirely determined by the two parameters $\eta_A$ and $\chi_A$,
 solutions of the linear system (\ref{linsys}). Thus, it depends only
 on the $G_n$'s values, which depend at turn on geometrical
 factors. Hence, the only relevant parameter is the ratio $Z_0/M$. We
 \begin{figure}
        \psfig{width=\columnwidth,height=8.5cm,file=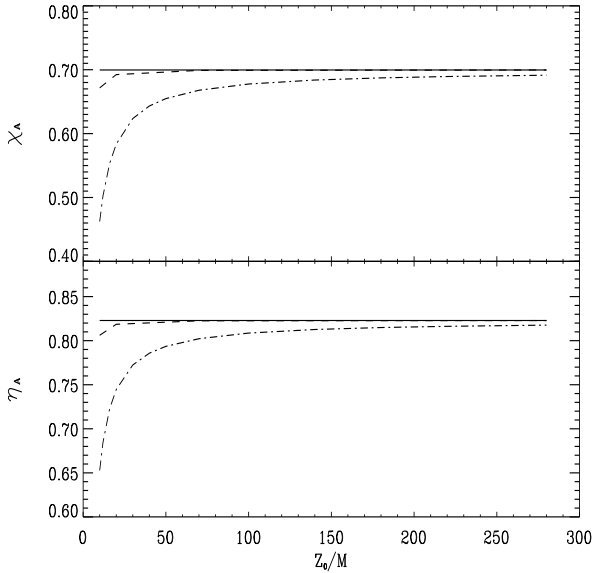}
        \caption[]{Parameters $\eta_A$ and $\chi_A$ versus
          $Z_0/M$. The Newtonian values without central hole are
          plotted in solid line and Newtonian values with central hole
          of radius $1.23M$ (the marginal stability radius corresponding
          to $a=0.998$) in dashed line. The Kerr values with $a=0.998$
          are plotted in dash-dotted line}\label{etachi}  
 \end{figure}
 \begin{figure}
        \psfig{width=\columnwidth,height=8.5cm,file=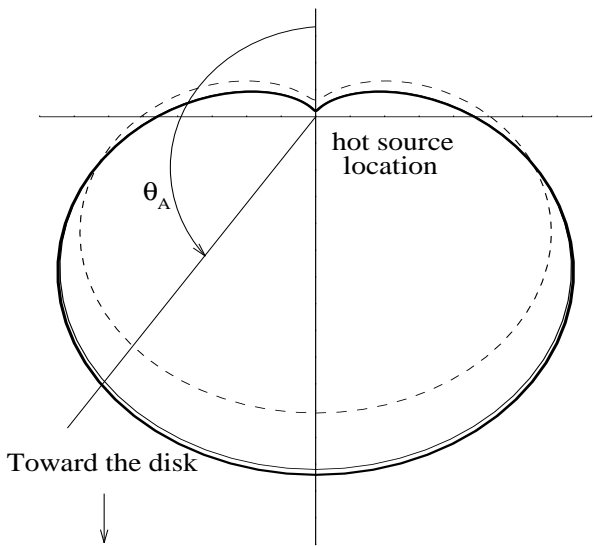}
        \caption[]{Polar plots of $\displaystyle\frac{dP}{d\Omega}$ for
        $Z_0/M=100$ 
        (solid line) and $Z_0/M=10$ (dashed line). The photon field at
        the hot source location is less anisotropic as the hot source is
        closer to the black hole. The bold line corresponds to the
        Newtonian case}
        \label{anisotropy}
 \end{figure}
 plot in Fig. \ref{etachi} the curves $\eta_A$ and $\chi_A$ as a
 function of $Z_0/M$ for $a=0.998$. The differences with the Newtonian
 case become 
 important for $Z_0/M\leq 50$, reaching about $30\%$ at $Z_0/M=10$.  
 The closer the source to the black hole is, the smaller $\eta_A$ and
 $\chi_A$ are. This corresponds to less anisotropic photon field. This is
 due to two effects: first the presence of a hole in the accretion disk
 inside the marginal stability radius; second the curvature of geodesics
 making the photons emitted near the black hole  
 arrive at larger angle than in the Newtonian case. As shown in Fig.
 \ref{etachi}, the first effect has a weaker influence than the second one.
 The polar plot of $\displaystyle\frac{dP}{d\Omega}$ is sketched in
 Fig. \ref{anisotropy} for $Z_0=100M$ and $Z_0=10M$.

\subsection{The radial temperature distribution}
\begin{figure}
        \psfig{width=\columnwidth,height=7cm,file=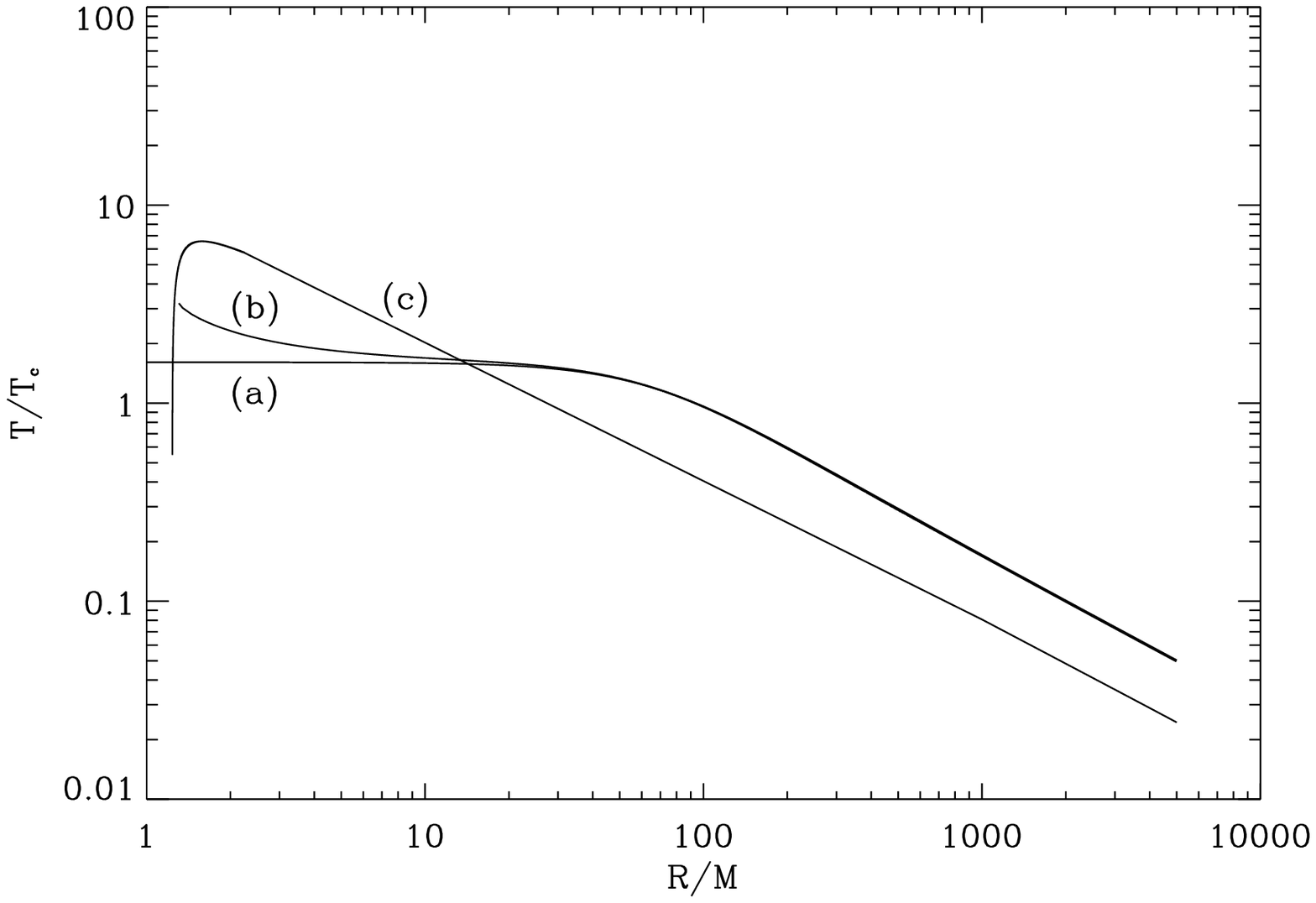}
        \caption[]{Effective temperature profile of the disk versus r in
          3 cases:\\ 
                            a) Our model in Newtonian metrics\\
                            b) Our model in Kerr metrics\\
                            c) Standard accretion disk\\
          We suppose the same total luminosity in each model
           }\label{temperature}
 \end{figure}

 We have plotted in Fig. \ref{temperature} the radial temperature
 distribution of three models: the Newtonian model of paper I, the
 present relativistic model with $a=0.998$ and $Z_0/M=70$, and the
 standard accretion disk model including relativistic effects
 (Novikov \& Thorne \cite{Nov73}) for the same total luminosity in each
 cases. The 
 temperature profile is 
 markedly different between the 
 two illumination models and the standard accretion disk one. At large
 distances, all models give the same asymptotic behavior $T\propto
 R^{-3/4}$ (cf. paper I). In the inner part of the disk ($R\leq Z_0$), in the
 illumination models, the temperature
 saturates around the characteristic temperature $T_c$. On the other hand,
 it keeps increasing in the accretion model, where the bulk of the
 gravitational energy is released at small radii.
 Thus, for rapidly rotating black hole, the main difference comes from the
 smaller marginal stability radius ($R_{ms}=1.23\,M$ for $a=0.998$, whereas
 $R_{ms}=6\,M$ for $a=0$). This increases a lot the accretion efficiency
 that goes from $\simeq 5.7\%$ for a Schwarzschild black hole ($a=0$), to
 $\simeq 42\%$ for a maximally rotating Kerr black hole. In the same time
 the central temperature reaches much higher values. As seen in Fig. 
 \ref{temperature}, these effects have much less influence in the
 illumination model. Indeed, the power radiated by the disk surface is
 essentially controlled by $\displaystyle\frac{dP}{d\Omega}$, which is
 approximately constant for $R\leq Z_0$ (i.e. $\theta_A\geq\pi /4$) as
 shown in Fig.  
 \ref{anisotropy}. So, the differences with the Newtonian model comes
 only from gravitational and Doppler shifts which are only appreciable
 for small radii ($R\leq 5M$). Thus, they concern only a small fraction of
 the emitting area at $T=T_c$, unless $Z_0$ is itself small enough. 

\subsection{Overall spectrum}
\subsubsection{Influence of the hot source's height}
 \begin{figure}
        \psfig{width=\columnwidth,height=7cm,file=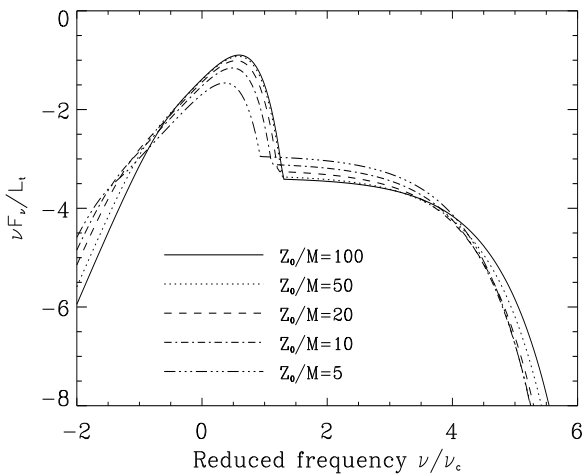}
        \caption[]{Differential power spectrum for different values of $Z_0$
          for the Kerr maximal case. We use
        reduced coordinates}\label{speczovar}
 \end{figure}
\begin{figure}
        \psfig{width=\columnwidth,height=7cm,file=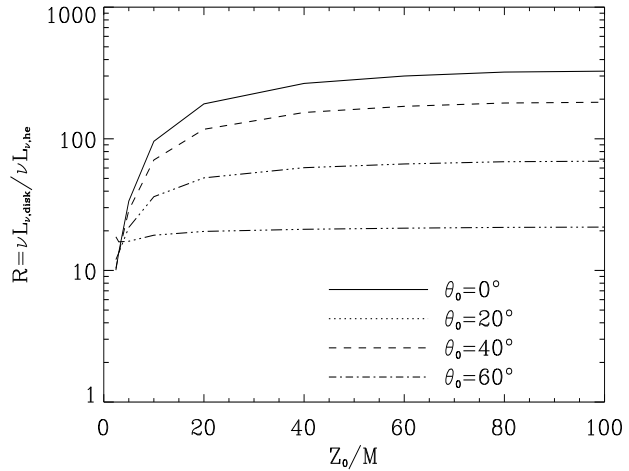}
        \caption[]{UV/X ratio versus $Z_0/M$ for different values of the
          inclination angle, in the Kerr maximal case.}\label{ratioX/UV(Z0)}
 \end{figure}

 Figure (\ref{speczovar}) shows the overall spectrum, in reduced units,
 predicted by the 
 model for different values of $Z_0/M$, for $L_t=~10^{45}\,erg.s^{-1}$,
 $M=5\,10^{6}\,M_{\sun}$ and $\theta_0=0^{\circ}$. The frequency shift 
 at both ends of the spectrum is due to the
 variations of the characteristic frequency $\nu_c$ with
 $Z_0$ (cf. Eq. (\ref{nuc})). The  
 relativistic effects themselves become important for values of $Z_0/M$
 smaller than about $50$. They produce a variation of intensity lowering
 the blue-bump and increasing the hard X-ray emission. The change in the
 UV range is due to the transverse Doppler effect between the rotating
 disk and the observer, producing a net redshift, the influence of this
 redshift being more important for small $Z_0/M$ as already explained in
 the last paragraph. In the X-ray range,
 the variation is due to the change of the parameters $\eta_A$ and
 $\chi_A$ when $Z_0/M$ decreases (cf. Fig. \ref{etachi}). The observed UV/X
 ratio can then be strongly altered by these effects. Quantitatively, the
 ratio between the maximum of the ``blue-bump'' and the X-ray plateau of
 our spectra,
 goes from $\simeq 300$ in the Newtonian case (or, equivalently, for high
 values of $Z_0/M$ in the Kerr metrics), to $\simeq 10$ for 
 $Z_0/M=3$ and $\theta_0=0^\circ$ in the Kerr maximal case, as shown in
 Fig. \ref{ratioX/UV(Z0)}. This ratio is highly dependent on the
 inclination angle $\theta_0$. By taking the maximum of the``blue-bump''
 ,which may be not observed, we evidently overestimate the UV/X ratio
 compared to the observations. It appears also that a small value of  
 $Z_0/M$ could explain the comparable UV and X-ray fluxes observed
 in few Seyfert galaxies (Perola et al. \cite{Per86}, Clavel et
 al. \cite{Cla92}). This behavior is clearly the opposite of what we would
 expect for a hot source whose emission is independent of the disk
 emission, and thus does not depend on $Z_0/M$. In such a case, the smaller
 the height of the hot source is, 
 the larger the bending effects on the ray of light emitted by the hot
 source are, increasing the illumination of the disk and thus increasing
 the UV/X ratio (Martocchia \& Matt \cite{Mar96}). It does not take
 into account the changes in the hot source emission due to the same
 bending effects and our model shows that, in this case, the global
 result is an increase of the X-ray flux toward the observer.\\

\subsubsection{Influence of the inclination angle}
 \begin{figure*}
        \psfig{file=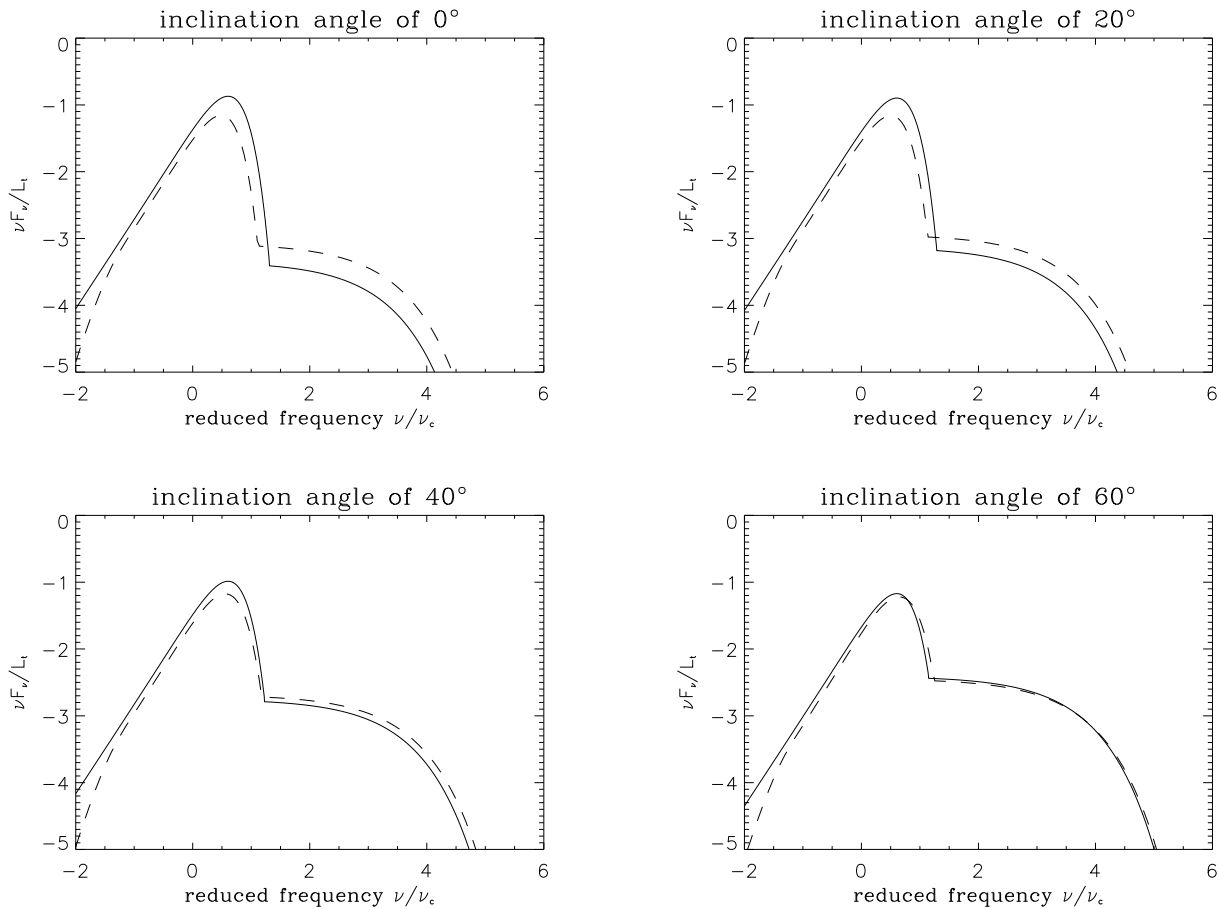}
        \caption[]{Differential power spectrum for different inclination
          angle, in the Newtonian (solid lines) and the Kerr maximal
          (dashed lines) cases for $Z_0/M=10$. We use reduced
          coordinates}\label{specplot} 
 \end{figure*}
 One can see on Fig. \ref{specplot} Newtonian and Kerr maximal
 spectra for different inclination angles for $Z_0/M=10$. For small
 inclination angles, the Kerr spectra are always weaker in UV and
 brighter in X-ray than the Newtonian ones. However, the difference tends
 to be less visible for the highest inclination angles. These results can be
 easily explained: in the X-ray range, as shown in
 Fig. \ref{anisotropy}, it is due to the decreasing of the relative
 difference of the angular distribution between Newtonian and Kerr
 metrics, as the inclination angle increases. But, for very small values
 of $Z_0/M$, the gravitationnal shift can be so high that the Kerr
 X-ray spectra appears weaker than the Newtonian one.
 In the UV band, the relativistic effects (the gravitational shift and the
 Doppler transverse effect) produce a net redshift in the
 face-on case ($\theta_0=0\degr$) compared to the Newtonian case. For
 higher inclination angle, the 
 redshifted radiation is compensated by the blueshifted one, coming from
 the part of the disk moving toward the observer.\\
 \noindent
 These effects are much less pronounced
 for high $Z_0/M$ values because the emission area is much larger, and
 thus is less affected by relativistic corrections.

 \section{Conclusion}
 We have studied the effects of general relativity on the spectrum emitted
 by our model of re-illumination of the accretion disk of Seyfert I
 galaxies by a relativistic plasma of leptons. This hot source could be the
 result of a strong shock between an abortive jet coming from the disk
 and the interstellar medium. It appears that,
 by opposition 
 with the standard model of accretion disk including relativistic effects
 (Sun \& Malkan \cite{Sun89}; Novikov \& Thorne \cite{Nov73}), there are
 few differences between 
 the Newtonian and relativistic case, unless the height of the hot source
 is small enough, i.e. $Z_0 \leq 50 M$.
 Indeed a region of the disk, of length scale of the order of the 
 height $Z_0$ of the hot source above the disk (and thus no disturb by the
 presence of the black-hole for large $Z_0$), is equally illuminated and
 thus predominates in the spectrum.\\
 \noindent
 In a future work, we will study the detailed structure of the hot
 source, taking into account the microphysical processes like pair
 production and particle acceleration. As mentioned in paper I, these
 processes could determine the unknown parameters (upper energy cut-off,
 spectral index, disk temperature) that are still free in the present
 theory.

\appendix
\section{Expression of $dS$}
\label{append1}
The elemental area $dS$ of a parallelogram defined by 2 vectors
$\vec{dl_1},\ \vec{dl_2}$ can be expressed, in curvilinear coordinates,
in a general form given by differential geometry. Locally, we can suppose
the space to be flat and characterized by 3 vectors $\vec{e_1},\ 
\vec{e_2},\ \vec{e_3}$. The projections of $dS$ on each plan
$(\vec{e_\alpha},\vec{e_\beta})$ are given by the antisymetrical tensor
$dS^{\alpha\beta}$:
 \beq
   dS^{\alpha\beta}=dl_1^\alpha dl_2^\beta -dl_1^\beta dl_2^\alpha
 \eeq
 In a 3 space, we rather use the vector $dS_{\alpha}$, dual of
 $dS_{\alpha\beta}$, defined by:
 \beq
   dS_{\alpha}=\frac{1}{2}\sqrt{\gamma}e_{\alpha\beta\gamma}dS^{\beta\gamma}
 \eeq
 $\gamma$ is the determinant of the space metric and
 $e_{\alpha\beta\gamma}$ is the Levi-Civita tensor. We can now obtain the
 expression of the surface $dS$ modulus of $\vec{dS_{\alpha}}$:
 \beqar
   dS &=& \sqrt{dS_{\alpha}dS^{\alpha}}\\ 
      &=& \sqrt{\gamma^{\alpha\beta}dS_{\alpha}dS_{\beta}}\\ 
      &=& \sqrt{\gamma\gamma^{\alpha\beta}
       \frac{1}{2}e_{\alpha\mu\nu}dS^{\mu\nu}
       \frac{1}{2}e_{\beta\mu'\nu'}dS^{\mu'\nu'}}
 \eeqar
 By setting the 4-tensor $G$ which coefficients are defined by:
 \beq
 \label{eqG}  
 G_{\mu\nu\mu'\nu'}=\gamma\gamma^{\alpha\beta}
 e_{\alpha\mu\nu}e_{\beta\mu'\nu'}
 \eeq
 we can re-write the expression of the elementary surface $dS$ (sum on
 repeated indices):
 \beq
   dS= \sqrt{G_{\mu\nu\mu'\nu'}dl_1^{\mu}dl_2^{\nu}dl_1^{\mu'}dl_2^{\nu'}}
 \eeq

\section{Kerr metric case. Expression of $dS$ in the plan $\theta=0$}
 In Kerr metric, the space metric tensor is diagonal
 \beqar
   \gamma_{\alpha\beta} &=& g_{\alpha\beta}-\frac{g_{0\alpha}g_{0\beta}}
   {g_{00}}\\ &=& 0 {\ \rm{if\ }}\alpha\neq\beta
 \eeqar
 To obtain the expression of the elementary surface $dS$ in the plan
 $\theta=0$ of the disk at radial coordinate $r$, we have to take
 $\alpha=\beta=\theta$ in  
 Eq. (\ref{eqG}). One gets then:
 \beq
  dS = \sqrt{\gamma_{rr}\gamma_{\varphi\varphi}}drd\varphi\\ \label{ds}
 \eeq
 As a matter of interest, the metric coefficients $g_{\alpha\beta}$ we
 use here, are calculated in the corresponding locally non-rotating frame
 $\Re$. In the 
 plane $\theta=0$ of the accretion disk, only the $g_{\varphi\varphi}$ metric
 coefficient differs 
 from the one of Boyer-Lindquist coordinate frame (Eq.\,(\ref{kerrmetric})).
 Thus, in frame $\Re$, $\gamma_{rr}$ and $\gamma_{\varphi\varphi}$ are
 equal to: 
 \beq
 \label{B4}
   \left\{
     \begin{array}{lll}
       \gamma_{rr} &=& \Sigma\Delta^{-1}\\
       & & \\
       \gamma_{\varphi\varphi} &=& e^{2\psi}
     \end{array}
   \right.
 \eeq
 A ring surface is obtain after integration of Eq. (\ref{ds}) with
 respect to $\varphi$, that is:
 \beq
  dS = 2\pi A^{1/2}\Delta^{-1/2}dr\\ 
 \eeq


\begin{thebibliography}{}
\bibitem[1972]{Bar72}
Bardeen J. M., Press W. H., Teukolsky S. A., 1972, ApJ 178, 347
\bibitem[1968]{Car68}
Carter B., 1968, Phys. Rew. 174, 1559
\bibitem[1992]{Cla92}
Clavel J., Nandra K., Makino F., et al., 1992, ApJ 393, 113 
\bibitem[1975]{Cun75}
Cunningham C. T., 1975, ApJ 202, 788
\bibitem[1973]{Cun73}
Cunningham C. T., Bardeen J. M., 1973, ApJ 183, 237
\bibitem[1991]{Ghi91}
Ghisellini G., George I. M., Fabian A. C., et al., 1991, MNRAS 248, 14
\bibitem[submitted]{Hen*}
Henri G., Petrucci P. O., A\&A, submitted
\bibitem[1996]{Mar96}
Martocchia A., Matt G., MNRAS, 282, L53
\bibitem[1973]{Nov73}
Novikov I. D., Thorne K. S., 1973. In C. DeWitt
and B. DeWitt (eds.) Black Holes. Gordon \& Breach, New York, p. 343
\bibitem[1986]{Per86}
Perola, G. C. , Piro, L. , Altamore, A., et al., 1986, ApJ 306, 508
\bibitem[1990]{Pou90}
Pounds K. A., Nandra K., Stewart G. C., et al., 1990,
Nat 344, 132 
\bibitem[1984]{Ree84}
Rees M, J., 1984, ARA\&A 22, 471
\bibitem[1973]{Sha73}
Shakura N. I., Sunyaev R. A., 1973, A\&A 24, 337
\bibitem[1989]{Sun89}
Sun W-H., Malkan M. A., 1989, ApJ 346, 68
\end{thebibliography}
\end{document}